\documentclass[runningheads]{llncs}
\usepackage{graphicx}
\usepackage[utf8]{inputenc}
\usepackage{enumitem}

\usepackage[colorlinks=true, linkcolor=blue, citecolor=blue, urlcolor=blue]{hyperref}

\usepackage{fancyhdr} 
\fancypagestyle{firststyle}
{
   \fancyhf{}
   \fancyfoot[C]{\scriptsize{Privacy and Identity Management. Data for Better Living: AI and Privacy: 14th IFIP WG 9.2, 9.6/11.7, 11.6/SIG 9.2.2 International Summer School, Windisch, Switzerland, August 19--23, 2019, Revised Selected Papers, pp. 343--358. The final publication is available at Springer via \url{https://doi.org/10.1007/978-3-030-42504-3_22}}}
}

\begin{document}
\title{Annotation-based Static Analysis for Personal Data Protection}

\author{Kalle Hjerppe\inst{1}\orcidID{0000-0002-3737-4669} \and
Jukka Ruohonen\inst{2}\orcidID{0000-0001-5147-3084} \and
Ville Leppänen\inst{2} \\
\email{kalle.hjerppe@geniem.com} \and
\email{\{juanruo,ville.leppanen\}@utu.fi}
}
\authorrunning{K. Hjerppe et al.}
%
\institute{Geniem Oy, Finland  \and
University of Turku, Finland}

\maketitle

\begin{abstract}
This paper elaborates the use of static source code analysis in the context of data protection. The topic is important for software engineering in order for software developers to improve the protection of personal data during software development. To this end, the paper proposes a design of annotating classes and functions that process personal data. The design serves two primary purposes: on one hand, it provides means for software developers to \textit{document} their intent; on the other hand, it furnishes tools for automatic \textit{detection} of potential violations. This dual rationale facilitates compliance with the General Data Protection Regulation (GDPR) and other emerging data protection and privacy regulations. In addition to a brief review of the state-of-the-art of static analysis in the data protection context and the design of the proposed analysis method, a concrete tool is presented to demonstrate a practical implementation for the Java programming language.


\end{abstract}

\section{Introduction}

\thispagestyle{firststyle} 

The famous General Data Protection Regulation (GDPR) in the European Union~\cite{gdpr} places various new requirements for software architectures as well as their design, development, and maintenance. Thus, this paper builds on previous work on eliciting requirements from the GDPR in the context of software architectures and their design \cite{re19paper}. While the previous work concentrated on high-level design, the present work takes a step down to the actual implementation of data protection solutions during software development. The approach presented belongs to the domain of static analysis of software source code.

Data protection is relevant for both organisational and software security. Numerous recent data breaches, such as the high-profile Equifax breach of 2017~\cite{equifax}, blatantly demonstrate how negligence has real consequences. The GDPR mandates organisations to invest in data protection to be able to process personal data securely and legally. For the present purposes, the juridical aspects can be also used to distinguish the concept of \textit{data protection} from the technical concept of (information) \textit{security}, which does not cover the legality of data processing. Furthermore,  the concepts of \textit{privacy} and data protection are often used synonymously, but the latter is important without the presence of the former. That is, data should be protected even in a context that does not respect~privacy.

The GDPR imposes various implicit requirements for software engineering, software development, and software architectures. Among these are general requirements related to concepts such as confidentiality, integrity, and availability. In several occasions, the GDPR also mentions \textit{appropriate} data protection measures (see Articles 5, 25, and 32). These measures are not binary-valued ``on/off requirements'' in their nature, however. The scale is continuous: the requirements can be improved with design patterns and different software development technologies. Static analysis is a good example about such technologies.

In essence, static analysis is a method of analysing a program (\textit{analysis}) without running it (\textit{static}). Static source code analysis can be applied to any software project with appropriate tools. It is also generally accepted as a ``\textit{security-by-design}'' best practice \cite{staticanalysissecurity}. Using tools to improve software quality during software development is also very much what the GDPR asks for. Therefore, static analysis naturally extends also to the present ``\textit{data-protection-by-design}'' context.

Given this background, this paper answers to two research questions (RQs):

\begin{itemize}

\item{RQ1: \textit{What parallels there are between those static analysis solutions designed for security and those that seek to improve privacy and data protection?}}

\item{RQ2: \textit{How the practical state-of-the-art of static analysis solutions can be applied in the context of the GDPR's data protection requirements?}}

\end{itemize}

The answer to the first research question is sought by analysing and categorising a few recent static analysis tools. There is a well-established literature on static analysis in general and using static analysis for improving security in particular (see, for instance, \cite{staticanalysissecurity} and \cite{staticanalysis}). There exists also some previous work on using static analysis for the GDPR's requirements~\cite{taintgdpranalysis,staticanalysisgdpr}. However, a synthesis is still lacking---in fact, only little has been written about static analysis in terms of the intersection between security and data protection. 

Building upon the categorisation presented (RQ1), the second research question is answered by presenting a design of a concrete tool for using static analysis to improve data protection. To briefly outline the background of this tool and its design, it suffices to note that common static source code analysis tools used to prevent (security) bugs analyse the abstract syntax tree (AST) of a software source code under inspection. The so-called \texttt{FindBugs} \cite{findbugs} static code analyser is a good example in this regard. This inspection allows the tools to make judgements of the logical content of a program. By implication, however, the tools are also limited to the logical content; they cannot analyse design, developer intent, or data semantics. For these reasons, the tool presented augments AST inspection with personal data annotations.

The structure of the paper's remainder follows the two research questions examined. In other words, Section~\ref{sec: background} evaluates the state of the art and categorises static analysis solutions for data protection (RQ1). The subsequent Section~\ref{sec: approach} elaborates the tool implemented for improving personal data processing and its protection (RQ2). Section~\ref{sec: discussion} discusses the implications and future directions.

\section{Background and related work}\label{sec: background}

A sensible starting point is that security in general is a \textit{prerequisite} for data protection, which, in turn, is a partial legal prerequisite for privacy. If data is not secured, it is also meaningless to discuss data protection and privacy. In other words, compromising security allows to compromise the other two concepts. 

\begin{figure}
  \includegraphics[width=\linewidth]{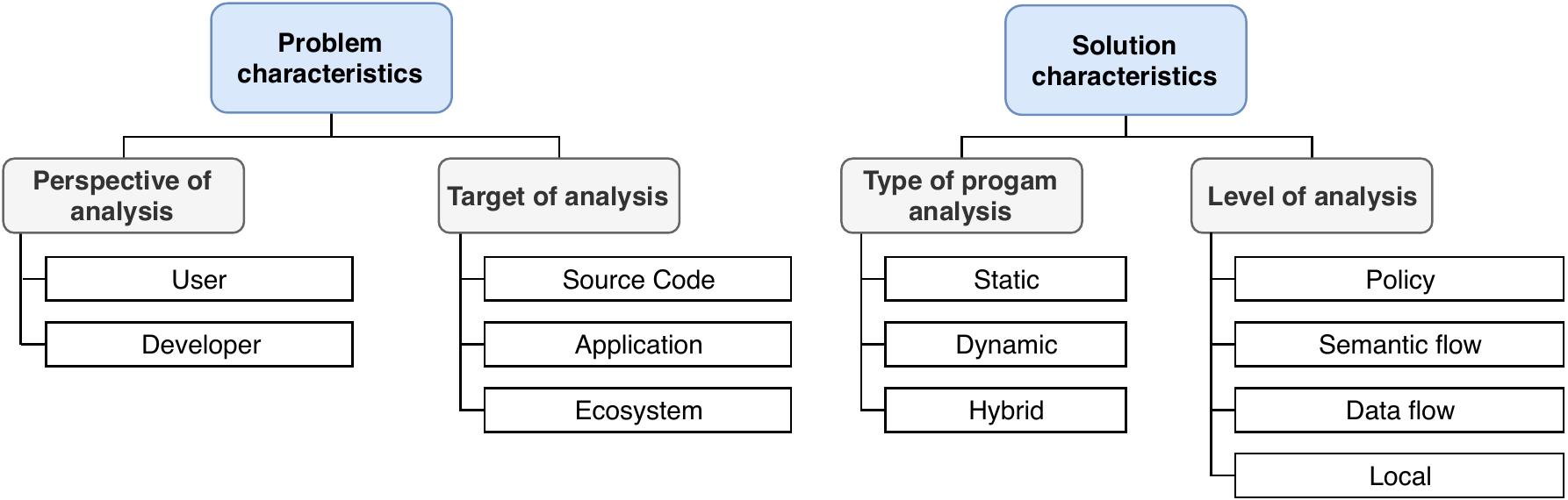}
\caption{Characteristics of existing approaches of analysis tools for data protection}
\label{fig:1}
\end{figure}

Given this assumption, a sensible next step for categorising the extensive literature on static analysis is the work of Sadeghi et al.~who consider both security and privacy (data protection) in the context of the Android operating system and its software ecosystem~\cite{sadeghi2016taxonomy}. They categorise different research approaches by considering the problems addressed (``what'') and the solutions to the problems (``how''). This fundamental taxonomy works also in the present context. Building upon their categories, different dimensions of data protection are thus illustrated in Fig.~\ref{fig:1}. By analysing these characteristics from the dual perspectives of data protection and security, it is possible to reason about differences as well as parallels. When these characteristics are combined, it is further possible to demonstrate a research gap. The following will briefly elaborate this gap.

\subsection{Categorisation by problem characteristics}

Threat modelling is a distinct characteristic of program analysis tools used for security assessments~\cite{sadeghi2016taxonomy}. In other words, an analysis tool seeks to address a particular threat or a set of threats. In terms of data protection, unintended information disclosure is the obvious threat---and a data breach a manifestation of the threat's realisation. For the present purposes, however, threat models are not sufficient for a categorisation of existing static analysis tools.

There appears to be a fundamental difference that separates static analysis approaches for data protection and privacy: whether an analysis is done from a \textit{perspective} of a user or a provider of the software the user is using. The user-oriented approaches typically focus on analysing software for potentially malicious or erroneous activity (see, for instance, \cite{185308} and \cite{yang2013appintent}). In contrast, the developer-oriented approaches typically focus on verifying and validating software. For example, Calcagno et al.~\cite{fbstaticanalysis} incorporate developer-oriented static analysis to software development for verifying memory safety. Obviously, it is possible to use a single approach from both perspectives; therefore, it is important to also consider the target of a static analysis tool.

In the present work, \textit{the target of analysis} refers to a particular software artifact under inspection. Traditional static analysis tools focus on source code. The perspective of analysis is on the developer side. In contrast, application-specific analysis is common for the user-oriented perspective already because the associated source code might not be public. The mapping between perspectives and targets is not rigid, however. It is possible to carry out developer-oriented application inspections, and so forth. For instance, the so-called \texttt{SCanDroid} tool may be used to analyse Android applications~\cite{fuchs2009scandroid}, yet the tool is still essentially developer-oriented since it provides means to certify security specifications.

The third target of analysis operates at the level of whole software ecosystems. By using different frameworks for dependency graphs, there have been various attempts to analyse software vulnerabilities at the level of whole ecosystems. A typical example is the \textit{npm} ecosystem and its dependencies for JavaScript libraries~\cite{npmanalysis,Zimmermann19}. Another example would be the software vulnerabilities in the packages stored to the Python's PyPI repository~\cite{Ruohonen18IWESEP}. However, according to the literature review conducted, no notable previous work has been done to extend these approaches toward data protection. 

\subsection{Categorisation by solution characteristics}

Turning to the how-question, a reasonable separation can be done in terms of the type and level of a software analysis solution. In terms of the \textit{type}, there are three common cases: static, dynamic, and hybrid. Each of these three categories have representatives also in the data protection and privacy contexts~\cite{arzt2014flowdroid,enck2014taintdroid,wang2019leakdoctor}. While each type has its merits~\cite{arzt2014flowdroid}, the paper's focus is strictly on static analysis.

The \textit{level} of analysis is the last but not the least dimension considered. The dimension builds on the noted premise that both data protection and privacy build upon security. This dimension is also the one through which many notable differences between different security and data protection (privacy) solutions differ. Four levels are considered: local, data flow, semantic flow, and policy. 

In terms of the local level, static analysis tools typically consider the syntax of the source code without much semantics and context. A good example would be so-called linters, which are reasonably well-suited for detecting simple syntax errors but also some more serious issues, such as Null pointer accesses. In contrast, the data flow, semantic flow, and policy levels seek to follow data through a software and make judgements based on this following. For instance, basic tools for finding structured query language (SQL) injections operate at the data flow level. Semantic flows add meaning to the data and its flows. For instance, the so-called \texttt{TaintDroid} tool \cite{enck2014taintdroid} labels sensitive data according to the source of it, such as sensors. Another recent approach is the so-called \texttt{DroidRista} tool~\cite{alzaidi2019droidrista}, which achieves high privacy leak detection rates also at the semantic flow level.

A policy-level analysis adds security and data protection (privacy) policies to the semantic flow of data in order to verify that a policy is followed. An example of a policy-level approach is the \texttt{MorphDroid} tool \cite{ferrara2015morphdroid}. However, there are also more comprehensive languages for formally specifying data protection and privacy policies. 
These levels of analysis are also an important aspect when considering differences between security-focused approaches and those designed for data protection and privacy. While there is a long history of using policies for security, many of the classical solutions do not consider semantics well. Whether it is mandatory access control mechanisms, group-based solutions, or capabilities, data is often seen uniformly as data worth protecting regardless whether the data is sensitive personal data or not. In contrast, recent formal languages specific to data protection and privacy, such as the so-called Layered Privacy Language (LPL) introduced by Gerl at al.~\cite{Gerl2018}, have been strongly motivated by the semantics of personal data. This language provides different constructs to reason about privacy policies. For the purposes of the tool soon elaborated, the LPL's \textit{Policy}, \textit{Purpose}, and \textit{Data} constructs are particularly useful.

The categories briefly described are useful for distinguishing different approaches and their underlining perspectives. For instance, an interesting comparison could be done regarding the different levels of analysis when the user-oriented and developer-focused approaches are used. In fact, it seems that a lot of the literature about the policy-level is tied to the user-oriented perspective; a typical goal is to verify that an application follows a privacy policy. If this generalisation is accepted, it could be also stated that a goal of a developer-oriented policy-level tool, in turn, would be to \textit{generate} the given policy from the source code. On the developer side, however, many current tools specific to data protection and privacy operate on the data flow and local levels of analysis. This observation provides the rationale for the tool presented; the goal is to move from the data flow level to the arguably more important policy level.

\subsection{Related work}

The high-level categorisation presented can be used to group many (if not most) static analysis approaches. To augment the categorisation, the approach proposed can be further explicitly compared with the following four previous works:

\begin{enumerate}

\item{Taint analysis can be done both statically and dynamically. Static taint analysis has the potential to solve the same issues as the approach of this paper. For instance, the work of Arzt et al. \cite{arzt2014flowdroid} finds privacy leaks in Android applications. Their approach is limited to programs for which data sensitivity can be deduced as coming from user input. Server-side taint analysis, in turn, is limited to the data flow level of analysis---unless an appropriate context for data sensitivity can be provided (such as the annotations presented). As a consequence, taint analysis tools are generally either for client-side analysis or they focus on security vulnerabilities in server-side applications.}

\item{Formal languages, coupled with a compiler, such as the one developed by Ramezanifarkhani et al. \cite{ramezanifarkhani2018secrecy}, possess the potential to make privacy policies statically verifiable. These languages have the potential to statically enforce data protection stronger than the lightweight analysis proposed. Although these would limit the usefulness of static analysis approaches, it should be emphasised that the approaches are not yet ready for production use.}

\item{The work of Myers et al.~\cite{myers2000protecting} shares a similar goal than the approach proposed: information flow control. They use labels in source code (like annotations, but as an extension of a language) to protect variables from improper reading or modification. While their decentralised end-to-end approach is valid, it is user-focused in contrast to the developer-oriented perspective pursued in this paper. Their design is also more pervasive, requiring a specific execution platform and an extension for the programming language.}

\item{A lightweight static analysis design similar to the approach proposed has been presented by Evans et al. \cite{evans2002improving}. They leverage comments (acting as annotations) in C code in order to detect security vulnerabilities. While their approach is also developer-oriented, the low-level focus differs from the present work. Furthermore, the approach neither recognises personal data nor focuses on data protection as such. That is, the principle of annotating relevant parts of code is the same, but the approach proposed considers documentation as a supplement to technical verification of code.}

\end{enumerate}

\section{Tool}\label{sec: approach}

The tool proposed seeks to improve the analysis level of source code analysis tools by focusing on the developer-oriented perspective. The underlying rationale is to analyse personal data in order to prevent accidental leaks of this data. In essence, the tool's goal is to help software developers by warning them about possible privacy leaks in source code during software development or immediately afterwards. The focus is thus in the implementation phase in a typical software development life cycle. This focus allows to catch design and implementation mistakes early on. In this developer-oriented context, \textit{personal data} simply means sensitive values in run-time memory. A typical example is a typed object of a class. Traditional source code analysis cannot know which objects contain personal data without some outside information. This constraint places them to the data flow level and limits their usefulness for data protection. 

The goal is also easy to justify with respect to regulations. Personal data could be processed in an unintended way as a result of a bug. Saving values of personal data into a log file would be a good example. With regards to the GDPR, such a bug may violate confidentiality and the need to know where all personal data is stored. Warning a developer from possible unintended processing with a short feedback loop should improve these and other related risks.

A brief further point can be made about performance requirements. Recently, Distefano et al. have discussed the issue of scaling static analysis solutions for large code bases \cite{distefano2019scaling}. Their main scaling properties are \textit{composionality} \cite{calcagno2009compositional,cousot2002modular} and \textit{abstraction} \cite{cousot1977abstract}. They define composionality so that a program analysis is composional when ``\textit{the analysis result of a composite program is defined in terms of the analysis results of its parts and a means of combining them}'' \cite[p.~70]{distefano2019scaling}. Abstraction, on the other hand, refers to considering only parts of the procedures that are relevant to the analysis, and discarding the rest. The gain in having these properties is the ability to parallelise the analysis, or doing incremental analysis only on parts of a code base. These properties are taken into account in the approach proposed---after all, scaling and performance are important aspects for any software development tool.

\subsection{Design of the tool}

The design of the tool has two layers. The first layer refers to the annotating of personal data objects in source code. This annotating is useful for raising the analysis level for source code analysis solutions in general. The second layer builds on a light source code analysis on the semantic flow level of analysis. This layer is constructed with two additional annotations to further specify a data processing context. The method demonstrates how a higher analysis level is reachable with relatively simple rules without a heavy analysis process. As will be discussed, the method scales also to larger problems.

The solution proposed uses the following three annotations:

\begin{itemize}
    \item{A1: \texttt{@PersonalData}}
    \item{A2: \texttt{@PersonalDataHandler}}
    \item{A3: \texttt{@PersonalDataEndpoint}}
\end{itemize}

In essence, the \texttt{@PersonalData} annotation (A1) should be used to document all classes containing personal data. This annotation gives the tool the necessary context to separate personal data types from other types. It also serves as a documentation of the source code. The other two annotations, \texttt{@PersonalDataHandler} (A2) and \texttt{@PersonalDataEndpoint} (A3), are used to further document the context within which personal data is processed. There are three specific rules (R) for this processing of personal data:

\begin{itemize}
\item R1: \textit{Data classes storing personal data are annotated with A1}.

\item R2: \textit{Any instance of an A1-annotated class used in a context not annotated with A2 is a violation}.

\item R3: \textit{In A2-annotated contexts, calling a non-A3 function outside the A2-context is a violation if an argument is A1-annotated.}
\end{itemize}

\begin{figure}
  \includegraphics[width=\linewidth]{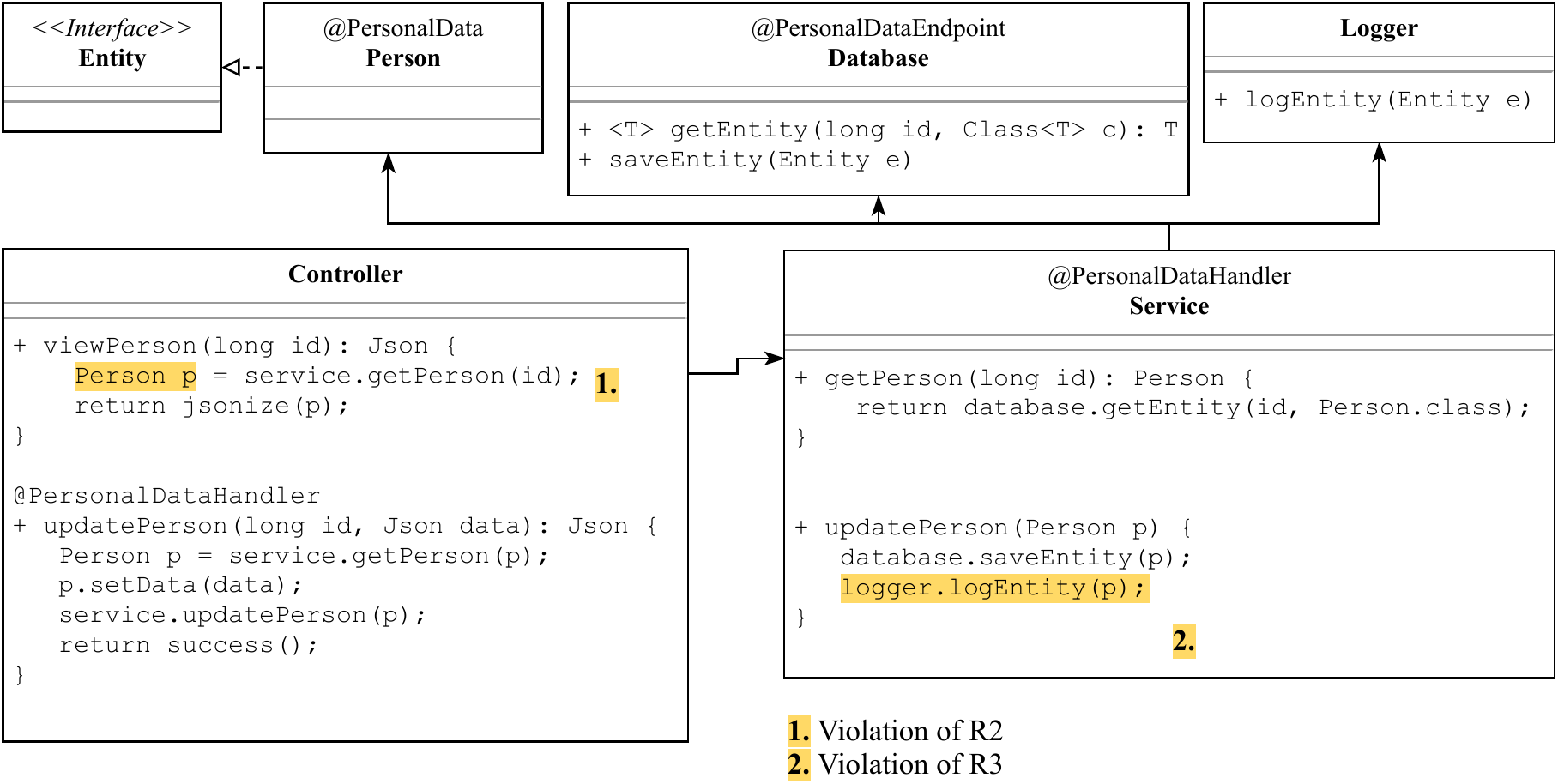}
\caption{Diagram and pseudo-code example of R2 and R3 violations.}
\label{fig:2}
\end{figure}

Rule R1 is the basis of the analysis, enabling the other two. R2 covers most data processing by requiring explicit handling of personal data to be documented. R3 allows establishing boundaries for personal data processing contexts within an application. A good example would be a generic database method for fetching a single entity. To further clarify, an A1-annotated class includes type variables for instances of generic classes and classes inheriting actual A1-annotated classes. Contexts annotated with A2 and A3, in turn, refer to functions or classes within which these are defined. The use of these three annotations is illustrated in Fig.~\ref{fig:2}. The illustration demonstrate a part of a class structure of an imaginary web application following the  classical \textit{model-view-controller}  design pattern. The illustration also demonstrate basic violations of the R2 and R3 rules. 

Three additional points are warranted about the design. First, in LPL's terms, A1-annotated classes correspond with (groups of) \textit{Data} elements. Such elements can be further mapped to distinct entry points, such as function-to-URL mappings of a web service. With such mappings and further annotations, it is possible to construct distinct execution trees that mimic LPL's \textit{Purpose}.

Second, the design adheres to the noted composionality requirement: a method under analysis only needs to know the annotations present in referred methods and classes, not whether they themselves would pass the analysis. By implication, the analysis can focus on units of code rather than the whole system---though, of course, to reach full coverage, eventually all of the code has to be analysed. When compared to more comprehensive designs, the design is also lightweight. Given that using static analysis for software development is perceived to have a significant cost~\cite{chatzieleftheriou2011test}, the lightweight design can ease adoption and shorten the feedback cycle for software developers. This practical point should not be undermined.

Third, in the terms of data flow analysis, the design can be briefly evaluated also in terms of so-called sensitivity properties~\cite{nielson2015principles}. The design discards path sensitivity and instead considers all possible branches of execution. Flow and context sensitivity are both needed to validate the rules described. Here, flow sensitive analysis implies being dependent on the order of instructions in the code, while context sensitivity comes from taking into account the calling context of method calls instead of analysing functions in a vacuum.

\subsection{Implementation}

The tool was implemented for the Java programming language using the Annotation Processing Tool (APT) functionality \cite{rocha2011annotations}. The implementation is also packaged and can be thus attached as a dependency for further projects. Command line build tools such as \textit{Maven}\footnote{https://maven.apache.org/} and integrated developer environments like \textit{IntelliJ IDEA}\footnote{https://www.jetbrains.com/idea/} allow adding annotation processors to projects' build processes. Therefore, the implementation can be also embedded to continuous integration environments. Extending IDEs to integrate further tools to enhance developers is a common practice (see \texttt{DebtFlag}, for instance \cite{holvitie2013debtflag}). The source code of the implementation is published\footnote{https://github.com/devgeniem/personaldataflow} under an open source license.

The implemented tool consists four classes: the three annotations discussed and a \texttt{PersonalDataAnnotationProcessor} (PDAP) class. The annotation processor inherits the APT's \texttt{AbstractProcessor} and implements \texttt{TaskListener} to hook into the Java compile process. This hooking allows to carry out the static analysis each time a software is compiled. A more heavyweight solution would not allow such a fast feedback loop. When compared to using only the \texttt{AbstractProsessor} model, which is limited to method signatures, making use of \texttt{JavacTask} allows traversing the entire AST of a program. Analysing the AST instead of the source code eliminates unnecessary noise from the analysis.

\begin{figure}
  \includegraphics[width=\linewidth]{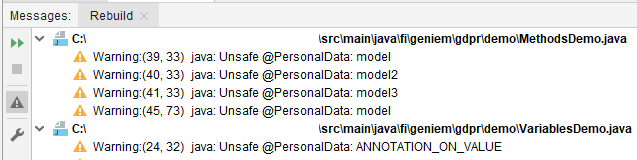}
\caption{Example of warnings produced by the analysis in \textit{IntelliJ IDEA}.}
\label{fig:3}
\end{figure}

After compilation of the target program, the PDAP class is notified by a \texttt{finished(TaskEvent task)} method call. This call provides the access to the compilation unit, and allows to traverse the AST with the \textit{visitor}~\cite{visitor} pattern. The result is a warning for each violation of R2 and R3, with a marker to the specific line of code. An example of presenting the violations is shown in Fig.~\ref{fig:3}.

The rules defined cover typical cases of processing personal data. The Java implementation passes conventional tests. However, there is one known issue: the tool does not cover representation exposure. For instance, a warning is not produced when extracting a primitive value from an A1-annotated object and then logging it. Improving the coverage of the tool for these situations should be investigated in further work. The current advise for this problem is to use specific classes for values of personal data rather than primitives. In the future, the coverage of the tool also improves as missing cases are added when discovered.

\section{Discussion}\label{sec: discussion}

The approach presented and the tool implemented improves data protection qualities of software systems processing personal data. As discussed by Chess and McGraw \cite{staticanalysissecurity}, the goal of static analysis for software security should be ``good enough'' coverage rather than guarantees. This objective underlines the developer-focused viewpoint to static analysis---to find as many bugs as possible. Privacy and data protection bugs do not mark an exception from this objective.

While the static analysis tool presented finds possible development errors, the \texttt{@PersonalData} annotations also document the nature of personal data in source code. If even more lightweight approach is desirable, a project might use only the \texttt{@PersonalData} annotations. When all three are used, processing of personal data is documented in source code rather extensively. Obviously, the downside is the increased effort for developers to annotate the code and maintain the annotations. While there is no single universal answer to a question about whether the quality increases are worth the effort, the annotations provide opportunities to develop further tools that increase the return on investment.

In other words, the annotations described enable the development of further tools, whether for code analysis, visualisation, or software analytics. Annotation-based reasoning is not strictly limited to source code analysis. It is possible to imagine an analysis of higher-level modules or even ecosystems in a similar fashion. In fact, a distant possibility would be to annotate all abstraction levels from source code; from source code to modules, from modules to systems, and from systems to systems of systems, each securing their own level of abstraction. The design and the rules would work similarly. Although automation would likely be difficult, AST could be replaced with a graph of modules, and so forth.

In terms of more immediate advances, two areas appear especially promising: (a) metrics for personal data prevalence and (b) generating LPL's \textit{Policies} from source code statically. In addition, (c) further work is required for better understanding what static analysis actually means for GDPR compliance. These three topics are discussed in what follows. 

Before continuing, however, three limitations can be briefly noted. First, the approach described is not granular and thus does not support separating singular paths in source code, unlike for example annotating distinct allowed execution paths in the code. This is a trade-off between developer effort and sensitivity of results. Second, the Java implementation is unable to find processing that goes beyond standard language usage, such as using reflection or type casts extensively. A developer must take this into account when relying on the results of the analysis. Last but not least: when compared to the whole scope of the GDPR, the results of this paper are only a small part of building systems with better data protection. Needless to say, the entire issue goes beyond technical measures altogether.

\subsection{Metrics for personal data prevalence}

Software metrics are commonly used for evaluating source code. The use cases range from quality control (since Boehm et al.~\cite{boehm1976quantitative}) to predicting change prone areas \cite{romano2011using}, for example. Metrics for software security can also be derived from source code \cite{chowdhury2008security}. However, to the best of our knowledge, mature metrics for personal data processing in source code are lacking. The approach of using annotations naturally extends to metrics for personal data and its processing.

A good topic would be the distribution of personal data across a system. While centralisation may not be optimal for security, reducing the distribution of personal data is a worthwhile goal in data protection terms~\cite{re19paper}. In fact, the centralisation of personal data to as few occurrences as possible is in spirit with the data minification principle of the GDPR. It can be applied from the architectural level also to the source code level. To facilitate this goal, developing metrics about personal data in code bases is relevant.

Even simple metrics (such as the share of \texttt{@PersonalData} classes to all classes) might be useful for better understanding the data protection aspects of a particularly large software architecture. Cohesion and coupling of different personal data processing call trees might also be valuable. In this respect, there is also a good literature base to build upon~\cite{MakelaLeppanen09}. With associated metrics, refactoring could be guided for more robust personal data processing.

Regardless of particular metrics, the annotations described enable automatic calculation of the values. This aspect also signifies the usefulness of the semantic flow analysis level: without knowledge of the personal data semantics, calculating the values would need to be done manually or supplied in an \textit{ad~hoc} fashion.

Any metrics from source code are naturally not metrics of \textit{privacy} itself, as argued by Danezis et al. \cite{danezis2015privacy}. However, focusing on detecting \textit{code smells} in this way is much easier to reason about than handling the abstract privacy concept as a whole. Source code metrics about data protection could point out especially sensitive areas of software architectures to focus efforts on, for example.

\subsection{Finding LPL Purposes}

Another promising next step would be to generate LPL's \textit{Policies} (or templates for the policies) from \texttt{@PersonalData} annotations coupled with web framework annotations. Common use cases for Java applications include server-side application programming interfaces (APIs) that essentially listen to different HTTP requests' URLs and execute a part of an application code for each. Using these in combination with the annotations, it should be possible to generate also policies.

Consider, for instance, the Spring framework\footnote{https://spring.io/} \texttt{@RequestMethod} annotation. This annotation is used to map service URLs to application \textit{entry points}. In essence, this mapping means that the application code itself does not handle web requests; the framework controls which method is called by which HTTP request. Therefore, the application source code can be viewed as distinct directed graphs of (possible) method calls, starting from the endpoint method. Each graph can be understood to cover one possible way to call the application. In addition, a conventional use case is to do an Object Relational Mapping (ORM) to map Java classes to database relations using another library. From the Java Persistence API specification, \texttt{@Entity} is a common way to implement this mapping.

By having an application built as described, the relevant entry points can be considered as LPL's \textit{Purposes}. This part is enabled trivially by the annotations: if an entry point method has a \texttt{@PersonalDataHandler} annotation, it is a \textit{Purpose}. Following from that, the source code and the execution paths can be traversed for finding uses of classes annotated with both \texttt{@Entity} and \texttt{@PersonalData}. These classes would then form LPL's \textit{Data} elements. The accessed \textit{Data} elements and the different \textit{Purposes} using them provide a good starting point for generating a \textit{Policy}. 
The same idea could be extended to analyse communication between modules through HTTP, for instance. This would require mapping the \textit{Purpose} of the first module to the \textit{Purpose*} of another module, and a way to analyse the APIs statically in both ends. They would then combine into a higher abstraction level \textit{Purpose}. An example would be to use Web Service definitions with corresponding annotations on both the service provider and the client.

A weakness is that a certain architecture is required for an application in order to derive the policies. The idea also breaks down in case all functionality is coupled to a single entry point. It should be also noted that the framework code, being imported from a library, would be outside of the scope of the compile time analysis presented in this paper, which is an advantage in this case. The analysis would then start from the entry point and cover the execution flow.

\subsection{Regarding the GDPR and static analysis}

It is necessary to point out the obvious: it is not possible to achieve compliance with the GDPR merely with just static analysis. While technical requirements can be derived from the regulation~\cite{re19paper}, the GDPR itself does not lay any exact technical requirements for compliance. Against this backdrop, Schneider~\cite{pbconstruction} argues that \textit{a posteriori} methods, such as static analysis conducted after software is designed, are not useful for achieving ``\textit{privacy-by-construction}''. That would be the ultimate goal of privacy engineering research. As he concludes, however, achieving ``privacy-by-construction'' is extremely difficult and expensive---if not impossible. Thus, in practice, other methods must suffice.

What the GDPR actually requires is ensuring appropriate personal data security with appropriate technical and organisational measures. Integrity and confidentiality are the underlying concepts (Article 5). Article 25 further qualifies these with the following remark: 

\textit{
\begin{quote}
``Taking into account the state of the art, the cost of implementation and the nature, scope, context and purposes of processing as well as the risks of varying likelihood and severity for rights and freedoms of natural persons posed by the processing [...]''
\end{quote}}

This quote would allow to argue that ``privacy-by-construction'' is a laudable goal but not a requirement. If the state of the art progresses to fully robust solutions, the goal may turn into a requirement, however. 

In the meantime, the usefulness of quality control methods such as static analysis seems apparent. Software quality is arguably also the essential aspect when contemplating about compliance. Many pragmatic industry practitioners fulfil legal requirements with minimum viable products and concentrate their efforts to business concerns. Such pragmatic reasoning tends to downplay the fact that quality control is a requirement in most software projects. Although not defined in exact terms, quality is also present in the GDPR. Therefore, the question of using static analysis is similar to a question of whether to use unit testing for improving software quality. Both questions are often cultural issues in software development organisations; some tools and approaches are preferred over others, and any method that increases costs is often debated.

It is impossible to say whether static analysis would have prevented any given data breach. It is also hard to evaluate whether a potential leak caught in analysis would not have been fixed otherwise. Since the GDPR does require appropriate data protection measures, investments to data protection tools may signal that an organisation was not \textit{negligent} even if a breach was to happen. While this ought to provide a motivation for engineers in itself, having lightweight tools well-integrated to the software development process lowers the threshold of adoption.

\section{Conclusion}

This paper studied static analysis for data protection. In addition to categorising state of the art tools, the paper provided a novel approach to improve software data protection qualities. The implementation of the presented tool is available to the public. The conclusions to the two research questions can be summarised as follows:

Regarding RQ1, three parallels were found between security and data protection focused static analysis tools: although (a) data protection analysis tools build upon security analysis tools, (b) threat models are less useful to categorise data protection tools. That said, the most important difference is that (c) data protection is more concerned about data semantics. These semantics imply a need for a higher \textit{level of analysis} for static analysis tools, from local level to the data flow level and from there to the semantic flow and policy levels.

Regarding RQ2, to demonstrate the applicability of static analysis for meeting the GDPR's data protection requirements, a novel design of a rule-based source code annotation method was presented alongside a concrete light-weight implementation. Although further work is required, the tool alone is sufficient for concluding that static analysis is applicable also in the GDPR context.

\bibliographystyle{splncs04}

\end{document}